# Prediction of many-electron wavefunctions using atomic potentials: extended basis sets and molecular dissociation

Jerry L. Whitten

Department of Chemistry
North Carolina State University
Raleigh, NC 27695 USA

email: whitten@ncsu.edu

**Abstract**

A one-electron Schrödinger equation based on special one-electron potentials for atoms is shown to exist that produces orbitals for an arbitrary molecule that are sufficiently accurate to be used without modification to construct single- and multi-determinant wavefunctions. The exact Hamiltonian is used to calculate the energy variationally and to generate configuration interaction expansions. Earlier work on equilibrium geometries is extended to larger basis sets and molecular dissociation. For a test set of molecules representing different bonding environments, a single set of invariant atomic potentials gives wavefunctions with energies that deviate from configuration interaction energies based on SCF orbitals by less than 0.04 eV per bond or valence electron pair. On a single diagonalization of the Fock matrix, the corresponding errors are reduced 0.01 eV. Atomization energies are also in good agreement with CI values based on canonical SCF orbitals. Configuration interaction applications to single bond dissociations of water and glycine, and multiple bond dissociations of ethylene and oxygen produce dissociation energy curves in close agreement with CI calculations based on canonical SCF orbitals for the entire range of internuclear distances.

## I. Introduction

A new method of constructing many-electron wavefunctions for molecules and extended systems was recently proposed and applied to a set of molecules in ground state equilibrium geometries.[1,2] A one-electron Schrödinger equation based on special one-electron potentials for atoms is shown to exist that produces orbitals for an arbitrary molecule that are sufficiently accurate to be used without modification to construct single- and multi-determinant wavefunctions. The exact Hamiltonian is used to calculate the energy variationally and to generate configuration interaction expansions.

The most interesting conclusion of the earlier work is the existence of a single potential associated with a given atom that works equally well in different bonding environments such as in single and multiple bonded systems and for bonds between atoms of different electronegativity. For example, the carbon potential is the same in methane, benzene, graphene and chlorophyll, and in other molecules. The purpose of the present work is to assess the accuracy of the molecular orbitals produced by these one-electron potentials for extended basis sets and to determine if the orbitals provide a suitably accurate basis that can be used directly in a many-electron method such as configuration interaction (CI). Doing so would eliminate the need for self-consistent-field (SCF) calculations prior to CI. If, however, one wishes to obtain a single-determinant SCF solution, the method provides an excellent initial field. There are many other ways to generate initial fields for SCF calculations and Lehtola has thoroughly analyzed various methods that have been proposed.[3] See also recent work on constructing initial fields[4], and earlier methods for constructing wavefunctions[5-13]

It now evident that the special one-electron potentials, called Vqc potentials in the earlier work,[1,2] have a more fundamental significance than originally envisioned. The potentials are a property of the atom and remain invariant in different molecules and in different bonding environments. Thus, the potentials plus antisymmetry determine how electrons fill into spatial regions around and between nuclei in a system. Energies for a set of molecules representing different bonding environments are calculated using single determinant and multi-reference CI wavefunctions. Some of the molecules are small enough to be treated easily by coupled cluster expansions methods and it would be worthwhile to investigate the convergence of CC expansions using Vqc molecular orbitals. The present work is limited to the use of multi-reference CI expansions. If the test applications are sufficiently accurate, as is found to be the case, new ways to approach extremely large systems and embedding formulations become possible. Orbitals of a system are now known in advance from the one-electron calculation and this in turn enables density and exchange fields and orthogonality for constrained portions of a system to be rigorously expressed. As noted in the Conclusions section, new computational strategies can be developed. In this paper we assess the accuracy of the method for larger basis sets and non-equilibrium geometries of molecules and discuss other implications.

Atomic potentials are optimized by considering a small set of molecules representing different bonding environments. Potentials derived from the superposition of 2 to 4 gaussian densities have been reported for H, C, N, O, and F, and oxides of Si, Ti, Fe, and Ni.[1,2] Details of the optimization procedure and the choice of reference molecules are described in the Appendix. An assertion that the orbitals are accurate and the potentials are simple and invariant in different systems would appear at first to be extremely unlikely given the complexity of the Fock operator in self-consistent-field calculations. However, it is now evident that there is a similarity in how atoms bond in systems even when single or multiple bonds are formed. This can be discovered by starting with a single-determinant wavefunction for a given basis and localizing the molecular orbitals about individual nuclei in the system. The resulting localized orbitals are similar, to within a unitary transformation, and correspond to *in situ* 1s, 2s, 2p … atomic orbitals mixed with functions from nearest neighbor and more distant atoms. The primary requirement of an orbital generating potential is to describe the mixing of basis functions to produce these localized orbitals.

The earlier papers addressed the determination of the potentials and the accuracy of single- and multi-determinant wavefunctions for a variety of systems. A few excited electronic states were also examined. Most of the previous studies involved a double-zeta basis and there remain important questions of whether the accuracy is maintained when extended basis sets are considered or when molecules are in non-equilibrium geometries or undergoing dissociation. The purpose of the present work is to address these questions. It is encouraging that the accuracy found for equilibrium geometries and simpler basis sets is found to be maintained.

## II. Theory

Consider a molecule or other system described by the Schrödinger equation

$$H_{exact}\psi = (\sum_i [-\tfrac{1}{2}\nabla_i^2 - \sum_q \frac{Z_q}{r_{qi}}] + \sum_{i<j} r_{ij}^{-1})\psi = (\sum_i [T_i + \sum_q V_{qi}] + \sum_{i<j} r_{ij}^{-1})\psi = E_{exact}\psi$$

with $N$ electrons and nuclei designated by $i$ and $q$, respectively, and associate with the system a modified Hamiltonian, $H^0$, that contains a one-particle potential for each nucleus, $v_{qi}$,

$$H^0\psi = (\sum_i h_i)\,\psi = (\sum_i [T_i + \sum_q (V_{qi} + v_{qi})])\,\psi = E\psi$$

where $\quad \psi = (N!)^{-1/2} \det(\chi_1(1)\chi_2(2)\chi_3(3)\ldots\chi_N(N)) \quad$ and $\quad \chi_i = \varphi_m(spin)$

Spatial orbitals are obtained by solving the one-electron eigenvalue problem $h_1\varphi_m = \varepsilon_m\varphi_m$.

One way to introduce one-electron potentials is to make use of rigorous bounds such as

$$<\varphi(1,2)\,|\,r_{12}^{-1}\,|\,\varphi(1,2)> \;\leq\; <\varphi(1,2)\,|-\tfrac{1}{2}\nabla_1^2 - \tfrac{1}{2}\nabla_2^2\,|\,\varphi(1,2)>^{1/2} <\varphi(1,2)\,|\,\varphi(1,2)>^{1/2}$$

for arbitrary $\varphi(1,2)$, or similar bounds containing one-electron potentials,[5] and apply the bounds to terms in $<\psi\,|\,H_{exact}\,|\,\psi>$. Such bounds on the total energy are unlikely to be accurate enough to be useful in complex systems, but the wavefunction obtained by minimizing the bound may be a useful starting point for variational calculations using the exact $H$.

Another way to proceed is to construct potentials that produce wavefunctions that closely match a variationally determined solution. This is the approach used in the present work. We start with one of the simplest definitions of a potential, one derived from densities centered at nuclei expanded as a linear combination of normalized spherical Gaussian functions, $\rho_a = (\frac{a}{\pi})^{3/2} \exp(-ar^2)$. For a single component (at nucleus $q$), the repulsive potential acting on an electron at position $i$ is $v_{qi} = \int |\vec{r_i} - \vec{r}|^{-1} \rho_a d\tau = 2\sqrt{\frac{a}{\pi}} r_i^{-1} \int_0^{r_i} \exp(-ar^2) dr$

Potentials are determined as described in Ref. 2 except for F where the optimization was improved. A Nelder-Mead simplex procedure[14] for optimizing densities and corresponding potentials is described in Refs. 1 and 2 and is summarized in the Appendix. The result of the optimization is a set of density parameters, $\{c_a, a\}$, for each atom $q$ where

$$\rho_q = \sum_a c_a \rho_a = \sum_a c_a (\frac{a}{\pi})^{3/2} \exp(-ar_q^2).$$

It is further assumed that the density is normalized to the nuclear charge, $Z_q = \sum_a c_a$, such that the nuclear attraction plus the repulsive potential is asymptotically zero for each nucleus.

The atomic potentials, $V_q$, referred to as Vqc potentials, are unique in the sense that they would correspond to the potentials found for a universe of molecules in which the deviation of individual molecules from the exact single-determinant solution is minimized. The important question of course is whether individual deviations are small enough for the potential to be useful and whether the accuracy depends on the basis set. If potentials exist that give acceptably small individual deviations, then a practical way to determine potentials would be to carry out optimizations on a series of molecules representing different bonding environments,[1,2] or as explored in this work for the F potential, to construct a super-molecule of widely separated individual molecules representing different types of bonding. The C and H potentials were obtained by optimization of only one molecule, $C_6H_6$. Next, given the C and H potentials, the N potential was optimized for $N_2C_4H_4$, and then the O potential for $NH_2CH_2\text{-}COOH$. If potentials exist that produce only a small deviation for each individual molecule from the exact, as is now known to be the case, it suggests considerable latitude in the choice of reference molecules used to determine potentials. For example, the different reference molecules used in Refs. 1 and 2 produce very similar results.

The result of the optimization is a set of atomic parameters reported in Table 1. The resulting potentials, including the nuclear attraction, are plotted in in Figure 1. Potentials are smoothly varying and thus applications should not be sensitive to the basis set except possibly for the interaction of short-range basis functions near the nucleus where the potential varies rapidly. The 1s atomic orbital is constrained to be doubly occupied to avoid possible inaccuracies at short-range. Molecular orbitals from the solution of the one-electron equation using these fixed atomic potentials are used directly in single-determinant wavefunctions and CI expansions.

Table 1. Atomic densities used to define potentials. Individual densities are normalized and coefficients sum to the nuclear charge $\rho_q = \sum_a c_a \rho_a = \sum_a c_a (\frac{a}{\pi})^{3/2} \exp(-ar_q^2)$

| | C | | | O | | |
|---|---|---|---|---|---|---|
| exp | 9.27800 | 0.33295 | 0.23450 | 5.80697 | 0.58286 | 0.11733 |
| coef | 2.40427 | 3.12627 | 0.46946 | 3.54700 | 3.50272 | 0.95027 |
| | H | | | F | | |
| exp | 0.31533 | 0.41983 | | 0.92495 | 14.64409 | 0.04312 |
| coef | 0.99952 | 0.00048 | | 5.59565 | 2.65663 | 0.74772 |
| | N | | | Si | | |
| exp | 3.70505 | 0.35327 | 0.09080 | 53.02198 | 0.43333 | 0.05449 |
| coef | 3.67864 | 2.75698 | 0.56437 | 6.29134 | 6.59128 | 1.11739 |

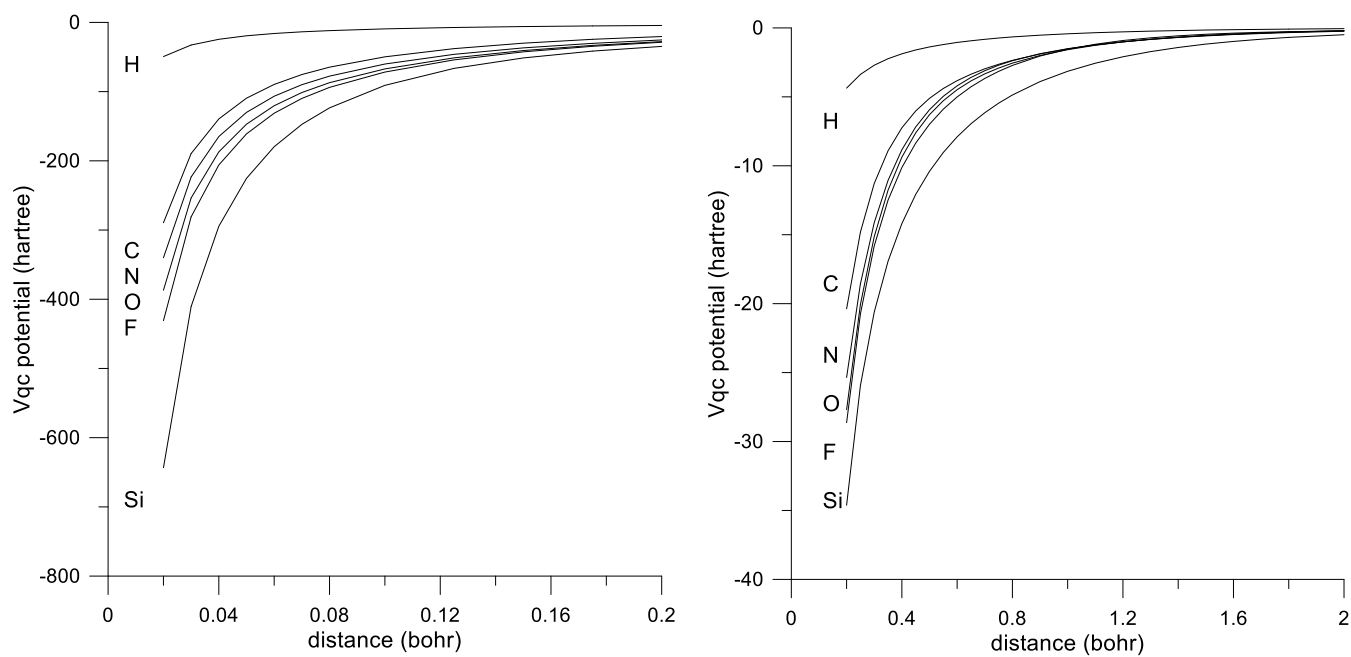

**Figure 1.** Plots of the total one-electron potential for atoms, $V_{qc} = v_q - \dfrac{Z_q}{r}$; short-range, left panel, longer range, right panel. Densities that generate potentials are normalized to the nuclear charge, and thus the potential asymptotically approaches zero for large distances.

Given these potentials, the one-electron eigenvalue problem is solved for each molecule of interest by minimization of $<\varphi_m | h_1 | \varphi_m >$ to determine basis function coefficients of $\varphi_m$. An antisymmetric wavefunction is then constructed as a single-determinant or a multi-determinant expansion.[15,16] All subsequent calculations use the exact Hamiltonian, explicitly including all electron repulsion interactions. The details of the CI procedure are described in the Appendix. Typically, for the larger molecules, ~$10^5$ configurations occur in the final CI expansion which can contain single through quadruple excitations from an initial representation of the state. Perturbation theory is used to estimate the contribution from configurations not explicitly included.[15-17] Of course, in the limit of a complete CI expansion, any linear transformation of the basis would give the same result, but for the larger systems, the present expansions are far from this limit.

**Accuracy for equilibrium geometries and triple-zeta basis sets**

We first address the accuracy for extended basis sets and then consider molecular dissociation of several single and multiple bonded systems. Single- and multi-determinant results for orbitals from the atomic potentials, reported in Table 2, demonstrate that the accuracy previously reported for smaller basis sets is maintained for the triple-seta basis and for even more extended basis sets used for ethylene, pyrazine, and the π-system of chlorophyll. The errors in total energy are equivalent to a maximum error of 0.04 eV per bond or valence electron pair for the multi-determinant expansions. Most errors are much smaller and the maximum occurs for large systems where it was not possible to include all orbitals in the CI.

Also included in the table are results from a single diagonalization of the Fock matrix constructed from the orbitals predicted by the Vqc potentials, denoted Vqc-diag, a calculation equivalent one SCF iteration using the Vqc orbital field. [18,19] The single-determinant description is found to be a Energies are in hartree units. A negative value for the CI error indicates that the Vqc value is lower than the value obtained using canonical SCF orbitals.

**Table 2. Comparison of energies for wavefunctions predicted by Vqc potentials with canonical SCF and CI calculations**

| molecule | total E[a] 1-det | total E ci | % error 1-det | % error ci | $\Delta E$[b] atoms | % error[b] | molecule | total E[a] 1-det | total E ci | % error 1-det | % error ci | $\Delta E$[b] atoms | % error[b] |
|---|---|---|---|---|---|---|---|---|---|---|---|---|---|
| **c6h6** | | | | | | | **glycine (nh2-ch2-cooh)** | | | | | | |
| exact | -230.6728 | -231.1714 | | | 1.7947 | | exact | -282.7876 | -283.2789 | | | 1.1232 | |
| vqc | -230.6615 | -231.1662 | 0.0049 | 0.0022 | 1.7911 | 0.20 | vqc | -282.7427 | -283.2587 | 0.0159 | 0.0071 | 1.1043 | 1.68 |
| vqc+diag | -230.6717 | -231.1709 | 0.0005 | 0.0002 | 1.7957 | -0.05 | vqc+diag | -282.7805 | -283.2786 | 0.0025 | 0.0001 | 1.1242 | -0.08 |
| **c4h4n2** | | | | | | | **nh2-ch2-coo(-) (removal of h+)** | | | | | | |
| exact | -262.6226 | -263.1475 | | | 1.3395 | | exact | -282.2127 | -282.6981 | | | 1.0391 | |
| vqc | -262.5823 | -263.1357 | 0.0153 | 0.0045 | 1.3291 | 0.78 | vqc | -282.1768 | -282.6805 | 0.0127 | 0.0062 | 1.0228 | 1.57 |
| vqc+diag | -262.6140 | -263.1457 | 0.0033 | 0.0007 | 1.3391 | 0.03 | vqc+diag | -282.2077 | -282.6981 | 0.0018 | 0.0000 | 1.0405 | -0.13 |
| **c4h4n2 (incl d)** | | | | | | | **c6h5-nh2** | | | | | | |
| exact | -262.6582 | -263.2504 | | | 1.4425 | | exact | -285.6964 | -286.2500 | | | 1.9315 | |
| vqc | -262.6247 | -263.2372 | 0.0128 | 0.0050 | 1.4306 | 0.82 | vqc | -285.6781 | -286.2432 | 0.0064 | 0.0024 | 1.9263 | 0.27 |
| vqc+diag | -262.6529 | -263.2489 | 0.0020 | 0.0006 | 1.4423 | 0.01 | vqc+diag | -285.6944 | -286.2523 | 0.0007 | -0.0008 | 1.9354 | -0.20 |
| **c5h5n** | | | | | | | **c5h5-cooh** | | | | | | |
| exact | -246.6513 | -247.1595 | | | 1.5672 | | exact | -380.3501 | -380.9458 | | | 1.8076 | |
| vqc | -246.6210 | -247.1493 | 0.0123 | 0.0041 | 1.5584 | 0.56 | vqc | -380.2990 | -380.9270 | 0.0134 | 0.0049 | 1.7910 | 0.92 |
| vqc+diag | -246.6470 | -247.1587 | 0.0018 | 0.0003 | 1.5678 | -0.04 | vqc+diag | -380.3425 | -380.9458 | 0.0020 | 0.0000 | 1.8098 | -0.12 |
| **c2h4** | | | | | | | **c6h5-cooh** | | | | | | |
| exact | -78.0241 | -78.2318 | | | 0.7794 | | exact | -418.2369 | -418.8300 | | | 1.4624 | |
| vqc | -78.0218 | -78.2322 | 0.0030 | -0.0005 | 0.7803 | -0.12 | vqc | -418.1789 | -418.8100 | 0.0139 | 0.0048 | 1.4448 | 1.20 |
| vqc+diag | -78.0239 | -78.2321 | 0.0003 | -0.0003 | 0.7801 | -0.10 | vqc+diag | -418.2273 | -418.8317 | 0.0023 | -0.0004 | 1.4665 | -0.28 |
| **ch4** | | | | | | | **fhco** | | | | | | |
| exact | -40.1895 | -40.3056 | | | 0.5859 | | exact | -212.7036 | -213.0527 | | | 0.4357 | |
| vqc | -40.1814 | -40.3047 | 0.0203 | 0.0023 | 0.5852 | 0.12 | vqc | -212.6460 | -213.0525 | 0.0271 | 0.0001 | 0.4363 | -0.14 |
| vqc+diag | -40.1882 | -40.3059 | 0.0033 | -0.0009 | 0.5865 | -0.10 | vqc+diag | -212.6927 | -213.0561 | 0.0051 | -0.0016 | 0.4399 | -0.96 |

| | c2h2 | | | | | | c2f2h2 | | | | | |
|---|---|---|---|---|---|---|---|---|---|---|---|---|
| exact | -76.8119 | -77.0181 | | | 0.5592 | | exact | -275.6796 | -276.1178 | | | 0.6453 | |
| vqc | -76.8034 | -77.0159 | 0.0112 | 0.0028 | 0.5575 | 0.30 | vqc | -275.5924 | -276.1058 | 0.0316 | 0.0044 | 0.6342 | 1.72 |
| vqc+diag | -76.8113 | -77.0166 | 0.0009 | 0.0019 | 0.5582 | 0.17 | vqc+diag | -275.6614 | -276.1172 | 0.0066 | 0.0002 | 0.6457 | -0.06 |
| | c2h6 | | | | | | c6h5-f | | | | | | |
| exact | -79.1705 | -79.3793 | | | 0.9333 | | exact | -329.5403 | -330.1173 | | | 1.7305 | |
| vqc | -79.1455 | -79.3777 | 0.0316 | 0.0020 | 0.9322 | 0.12 | vqc | -329.4976 | -330.0940 | 0.0130 | 0.0071 | 1.7090 | 1.25 |
| vqc+diag | -79.1657 | -79.3791 | 0.0061 | 0.0002 | 0.9336 | -0.04 | vqc+diag | -329.5347 | -330.1154 | 0.0017 | 0.0006 | 1.7304 | 0.01 |
| | h2o | | | | | | graphene model (c24h12) | | | | | | |
| exact | -76.0185 | -76.1628 | | | 0.2886 | | exact | -915.7804 | -916.1172 | | | 4.5716 | |
| vqc | -76.0131 | -76.1624 | 0.0071 | 0.0006 | 0.2884 | 0.04 | vqc | -915.7267 | -916.1113 | 0.0059 | 0.0006 | 4.5717 | 0.00 |
| vqc+diag | -76.0177 | -76.1629 | 0.0010 | -0.0001 | 0.2890 | -0.14 | vqc+diag | -915.7747 | -916.1123 | 0.0006 | 0.0005 | 4.5727 | -0.02 |
| | h2co | | | | | | chlorin (c20n4h16) | | | | | | |
| exact | -113.8444 | -114.0818 | | | 0.4749 | | exact | -984.2629 | -984.5745 | | | 4.1923 | |
| vqc | -113.8229 | -114.0826 | 0.0189 | -0.0007 | 0.4763 | -0.30 | vqc | -984.1589 | -984.5295 | 0.0106 | 0.0046 | 4.1529 | 0.94 |
| vqc+diag | -113.8395 | -114.0833 | 0.0043 | -0.0013 | 0.4769 | -0.44 | vqc+diag | -984.2492 | -984.5743 | 0.0014 | 0.0000 | 4.1977 | -0.13 |
| | co | | | | | | ethylene (58bf) | | | | | | |
| exact | -112.7099 | -112.9357 | | | 0.3223 | | exact | -78.0636 | -78.3681 | | | 0.9157 | |
| vqc | -112.6535 | -112.9349 | 0.0500 | 0.0008 | 0.3220 | 0.08 | vqc | -78.0546 | -78.3692 | 0.0116 | -0.0014 | 0.9173 | -0.18 |
| vqc+diag | -112.7068 | -112.9366 | 0.0028 | -0.0008 | 0.3238 | -0.47 | vqc+diag | -78.0630 | -78.3681 | 0.0008 | 0.0000 | 0.9162 | -0.06 |
| | nc4h5 | | | | | | si6o12 | | | | | | |
| exact | -208.7715 | -209.2197 | | | 1.3600 | | exact | -2631.0042 | -2631.8024 | | | 0.9459 | |
| vqc | -208.7535 | -209.2130 | 0.0086 | 0.0032 | 1.3545 | 0.41 | vqc-oxide | -2630.7764 | -2631.7818 | 0.0087 | 0.0044 | 0.9253 | 2.18 |
| vqc+diag | -208.7677 | -209.2186 | 0.0018 | 0.0005 | 1.3601 | -0.01 | vqc+diag | -2630.9682 | -2631.8132 | 0.0014 | -0.0011 | 0.9567 | -1.14 |
| | nc4h4 | | | | | | chlorophyll-a | | | | | | |
| exact | -208.1259 | -208.5659 | | | 1.2031 | | exact | -2369.4168 | -2369.5728 | | | 8.9189 | |
| vqc | -208.0949 | -208.5593 | 0.0149 | 0.0032 | 1.1975 | 0.46 | vqc | -2369.1554 | -2369.3706 | 0.0110 | 0.0085 | 8.7294 | 2.13 |
| vqc+diag | -208.1227 | -208.5654 | 0.0015 | 0.0003 | 1.2037 | -0.05 | vqc+diag | -2369.3642 | -2369.5225 | 0.0022 | 0.0021 | 8.8813 | 0.42 |

b ΔE = E(sum of atoms) - E(molecule). A negative value for the percent atomization error occurs when the CI calculation using Vqc orbitals describes the molecule more accurately (lower E) than canonical SCF orbitals.

greatly improved in all systems investigated and this usually leads to a lower CI total energy. The maximum error in Table 2. for the improved treatment is reduced to 0.01 eV per bond or valence electron pair. Thus, the diagonalization is very effective for those cases in which not all molecular orbitals are included in the CI. Another useful comparison would be to compute the energies required to decompose molecules fully into constituent atoms. Calculated atomization energies at the CI level, $(\sum_Q^{atoms} E_Q) - E_{molecule}$, are reported in Table 2. The agreement of calculations based on Vqc orbitals and Vqc+diag orbitals with the atomization energy obtained from CI calculations using canonical SCF orbitals is found to be quite good in all cases and exceptionally good for the Vqc+diag procedure.

**Molecular Dissociation**

Since the one-electron Vqc potentials were obtained by considering molecules in their equilibrium geometries, it is important to understand if the accuracy is maintained for non-equilibrium geometries. A sensitive test is the dissociation of molecules. To probe the question, the following dissociations are considered:

1) simple single bond, $H_2O \rightarrow OH + H$

2) complex single bond, $H_2N$-$H_2C$-$COOH \rightarrow H_2N$-$H_2C + COOH$

3) double bond, $H_2C$=$CH_2 \rightarrow H_2C + CH_2$ (CH2 in high spin, S=1 state)

4) triplet state multiple bond, $O_2 \rightarrow O + O$ ( S=1 state for O and O2 )

Potential energy curves are calculated at the CI level for a double zeta basis, using the same Vqc potential for all internuclear distances and results are compared with canonical SCF-CI calculations using the same basis. The Vqc calculations are at the simplest level where the molecular orbitals predicted by the Vqc potentials are used directly in the CI, i.e., with no intervening Fock matrix diagonalization. The exact H is used to evaluate integrals over molecular orbitals and matrix elements; thus, the energy is variational at each distance. Potential curves are depicted in Figure 1. In each graph, the solid line is a cubic spline fit to the canonical SCF-CI energies and the superimposed symbols (+) are energies obtained using Vqc molecular orbitals. The agreement is excellent over the entire range of distances. For glycine, an additional CI calculation based on Vqc-diag orbitals shows slightly better bonding because the Vqc virtual orbitals are better suited for correlation than those of the canonical SCF solution.

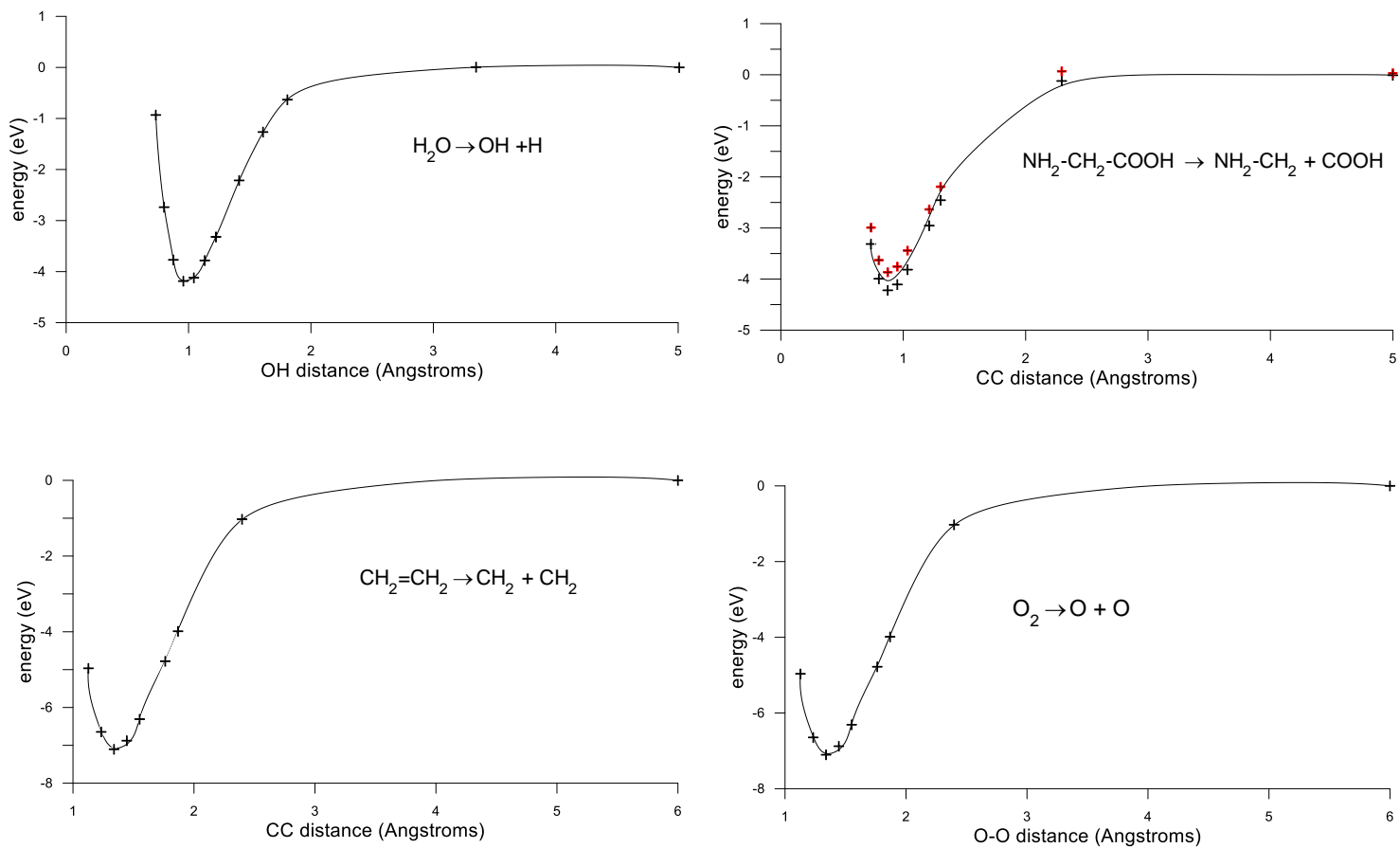

**Fig. 2. Dissociation of single and multiply bonded molecules.** The black crosses denote the energies obtained from CI calculations using Vqc predicted molecular orbitals. The solid line is a cubic spline fit to the canonical SCF-CI energies. For glycine, the red symbol is for a CI calculation based on Vqc-diag orbitals.

Another sensitive test of the potentials would be to minimize the Vqc energy, $E$, of a molecule by variation of the basis functions in $\varphi_m$ working only with the one-electron equation which does not contain electron repulsion explicitly,

$$(-\tfrac{1}{2}\nabla^2 + \sum_q v_q)\varphi_m = \varepsilon_m \varphi_m$$

$$E = \sum_m \varepsilon_m$$

In these calculations, 1s, 2s, and 2p basis functions which are atomic Hartree-Fock orbitals, are kept unchanged and only the extra longer-range basis functions are optimized by allowing exponents and coefficients to vary. The result of the calculation is a new basis for the molecule. The new basis is compared with the initial basis by performing canonical SCF calculations for each basis. Results for four tests are as follows:

|  | total energy (hartree) | |
| --- | --- | --- |
|  | initial basis | new basis |
| glycine | -282.7387 | -282.7771 |
| pyrazine | -262.5793 | -262.6151 |
| c6h5-cooh | -418.1783 | -418.2257 |
| chlorophyll-a | -2369.4168 | -2369.4249 |

The energy is lower for each system using the basis from the Vqc energy minimization. However, if shorter range (larger exponent) basis function components are also varied, the energies do not improve. This is a further indication of the point made earlier about short distances where the Vqc potentials, in Figure 1, are changing rapidly and may be insufficiently accurate to allow basis optimization using the one-electron equation.

**Conclusions**

This work establishes the existence of a special one-electron potential associated with a given atom that can be used to predict accurate molecular orbitals for a molecule. Potentials are reported for H, C, N, O and F and for Si in an oxide. The atomic potentials are a property of the atom and remain invariant in different molecules and in different bonding environments such as single-, double-and triple bonded systems. Only the wavefunction is predicted by the one-electron potential and the exact Hamiltonian including electron repulsion is required to determine the energy of single- and multi-determinant wavefunctions. The method is variational.

Earlier work examined double zeta basis sets and the present work shows that the accuracy is maintained for triple-zeta and larger basis sets. It is also shown that potential energy curves for several single- and multiple- bond dissociations are accurately described using the same one-electron potentials that were developed for equilibrium geometries.

In summary, the existence of invariant potentials suggests a similarity in how atoms bond in systems even when single or multiple bonds are formed. This can also be discovered by starting with a single-determinant wavefunction for a given basis and localizing the molecular orbitals about individual nuclei in a system. The resulting localized orbitals are similar, to within a unitary transformation, and correspond to *in situ* 1s, 2s, 2p … atomic orbitals mixed with functions from nearest neighbor and more distant atoms. The primary requirement of an orbital generating potential is to describe the mixing of basis functions to produce these localized orbitals. It is not surprising that simple spherical potentials contain sufficient flexibility to provide a good solution for atoms, but it is surprising that the mixing of basis functions on different nuclei is also well described by the same potentials and that different molecules can be reasonably well described by invariant potentials. We emphasize that only the wavefunction is predicted and the exact Hamiltonian, including electron repulsion, is required to determine the energy of single- and multi-determinant wavefunctions variationally.

If predicted orbitals are accurate enough to be used directly in correlated wavefunction constructions such as configuration interaction and if the atomic potentials are invariant there is an opportunity for constructing many-electron wavefunctions in a simple, but powerful way. Specifically:

1) States can be formulated precisely as eigenfunctions of spin, and ground and excited states and molecular dissociation treated correctly.
2) The self-consistent-field step and associated convergence problems are eliminated.
3) Orbitals are more efficient for CI expansions than virtual orbitals of a canonical SCF solution.

4) Small defects in molecular orbitals can be accounted for by single excitations in the configuration interaction expansion, or, if a better single determinant wavefunction is desirable, a single Fock matrix diagonalization can be performed.

5) If molecular orbitals predicted by the one-electron equation can be used directly in a configuration interaction expansion, it is unnecessary to save integrals over basis functions. Since molecular orbitals are known in advance from the one-electron solution, when the exact Hamiltonian is introduced, electron repulsion integrals over molecular orbitals can be calculated at the point integrals over basis functions are calculated.

**Appendix**

**Optimization of densities**

To construct average potentials designed to be transferable, we have found it better to consider molecules with complex bonding environments, rather than atoms or simple molecules, so that interactions are averaged over different environments. The present Vqc potentials for H, C, N and O are based on only three molecules: $C_6H_6$, $N_2C_4H_4$, and $H_2NCH_2$-COOH. The three systems were treated successively, keeping atomic parameters determined for preceding molecules invariant as described in Refs. 1 and 2  A new procedure was explored for F in which a supermolecule $C_6H_5$-F + $C_2F_2H_2$ + HFCO was defined as the target molecule for optimization of the Vqc potential for F. Such an approach can lead to slight improvements of the C, N and O potentials, but the potentials used for these atoms are from Ref. 2.

Simplex procedure:

1) Initial exponents and coefficients are specified as parameters for the constituent atomic densities (e.g., for benzene, exponents and coefficients for C and H densities). Suppose the parameter values are $w_1, w_2, w_3 \ldots w_p$.

2) The resulting one-electron eigenvalue problem is solved to determine energies and coefficients of basis functions in molecular orbitals, $\{\varepsilon_m, \varphi_m\}$. The lowest energy N spin orbitals are occupied.

3) A single determinant wavefunction is constructed from the predicted orbitals and its energy is evaluated using the exact Hamiltonian, a step that requires all electron repulsion integrals. The energy, $<\psi | H_{exact} | \psi>$, is a function of the parameters, $E(w_1, w_2, w_3 \ldots w_n)$.

4) Based on the value of $E$ and the current set of parameters, new parameters are selected and the process is repeated until $<\psi|H_{exact}|\psi>$ is minimized. The Nelder-Mead simplex procedure is a convenient way to accomplish this since the selection of new parameter values depends only on $E$ and the history of its variation with prior choices of parameters.[14]

The result of the optimization procedure is a set of density parameters, $\{c_a, a\}$, for each atom q in the molecule being considered, $\rho_q = \sum_a c_a \rho_a = \sum_a c_a (\frac{a}{\pi})^{3/2} \exp(-ar_q^2)$.

**Basis set**

The basis for each atom is a near Hartree Fock set of atomic orbitals plus extra one-component s- and p-type functions consisting of the two smaller exponent components of the atomic orbital. We refer to this as a triple-zeta basis. Orbitals are expanded as linear combinations of Gaussian functions: 1s(10), 2s(5), 2p(5), 2s′(2), 2p′(2), for C,N,O , 2p(6) for F, and 1s(4), s(1) for H where the number of Gaussian functions in each orbital is indicated in parentheses. The larger basis set used in the extended treatment of ethylene and glycine contains d- functions and chlorophyll contains an additional p-orbital in the π-system. No core potentials were used in the present calculations so that the predictive capability of the method could be fully tested.

**Configuration interaction**

All calculations are carried out for the full electrostatic Hamiltonian of the system

$$H = \sum_i^N [-\tfrac{1}{2}\nabla_i^2 + \sum_k^Q -\frac{Z_k}{r_{ik}}] + \sum_{i<j}^N r_{ij}^{-1}$$

A single-determinant self-consistent-field (SCF) solution is obtained initially for each state of interest. Configuration interaction wavefunctions are constructed by multi-reference expansions,[15-17]

$$\Psi = \sum_k c_k (N!)^{-1/2} det(\chi_1^k \chi_2^k ... \chi_N^k) = \sum_k c_k \Phi_k$$

In all of the applications, the entire set of SCF orbitals is used to define the CI active space. Single and double excitations from the single determinant SCF wavefunction, $\Phi_r$, create a small CI expansion, $\Psi'_r$,

$$\Psi'_r = \Phi_r + \sum_{ijkl} \lambda_{ijkl} \Gamma_{ij \to kl} \Phi_r = \sum_m c_m \Phi_m$$

The configurations $\Phi_m$, are retained if the interaction with $\Phi_r$ satisfies a relatively large threshold condition

$$\frac{|\langle \Phi_m | H | \Phi_r \rangle|^2}{|E_m - E_r|} > 10^{-4} \text{ a.u.}$$

The description is then refined by generating a large CI expansion, $\Psi_r$ by single and double excitations from all important members of $\Psi'_r$ to obtain

$$\Psi_r = \Psi'_r + \sum_m \left[ \sum_{ik} \lambda_{ikm} \Gamma_{i \to k} \Phi_m + \sum_{ijkl} \lambda_{ijklm} \Gamma_{ij \to kl} \Phi_m \right]$$

where $\Phi_m$ is a member of $\Psi'_r$ with coefficient $> 0.02$. We refer to this expansion as a multi-reference CI. The additional configurations are generated by identifying and retaining all configurations, $\Phi_m$, that interact with $\Psi'_r$ such that

$$\frac{|\langle \Phi_m | H | \Psi'_r \rangle|^2}{|E_m - E_r|} > 1 \times 10^{-6} \text{ a.u.}$$

Approximately $10^5$ configurations occur in the final CI expansion, and the expansion can contain single through quadruple excitations from an initial representation of the state $\Phi_r$.

**Acknowledgment**

A collaboration with Dr. Fariba Nazari on metal-metal bonding in transition metal systems is gratefully acknowledged as are helpful discussions with Professor Mike Whangbo.